\documentstyle[10pt,epsf]{article}
\epsfverbosetrue
\def\as{\alpha_{\rm s}}
\def\O{{\cal O}}
\def\ylab{y_{\,\rm lab}}
\def\ycms{y_{\rm CM}}
\def\lsim{\mathrel{\mathop
  {\hbox{\lower0.5ex\hbox{$\sim$}\kern-0.8em\lower-0.7ex\hbox{$<$}}}}}
\def\gsim{\mathrel{\mathop
  {\hbox{\lower0.5ex\hbox{$\sim$}\kern-0.8em\lower-0.7ex\hbox{$>$}}}}}
\def\to{\rightarrow}
\def\ms{$\overline{\rm MS}$}
\def\pT{p_{\rm T}}
\begin{document}
%
\begin{titlepage}
\noindent
DESY 93-034  \hfill  ISSN 0418-9833     \\
March 93\hfill                          \\[9ex]
\begin{center}
{\Large \bf Inclusive particle production at HERA: }    \\[1.6ex]
{\large \bf Higher-order QCD corrections           }    \\[1.3ex]
{\large \bf to the resolved quasi-real photon
            contribution                           }    \\[11ex]

{\large F.M.\ Borzumati                             }    \\[1ex]
{\large B.A.\ Kniehl                                }    \\[1ex]
{\large G.\ Kramer                                  }    \\[1.5ex]
{\it II.\ Institut f\"ur Theoretische Physik
\footnote{Supported by the Bundesministerium f\"ur Forschung und
 Technologie, 05 5HH 91P(8), Bonn, FRG.   } }       \\
{\it Universit\"at Hamburg, 2000 Hamburg 50, Germany}      \\[21ex]

{\large \bf Abstract}
\end{center}
\begin{quotation}
We calculate in next-to-leading order inclusive cross sections of
single-particle production via resolved photons in $ep$ collisions at
HERA. Transverse-momentum and rapidity distributions are presented and
the scale dependence is studied. The results are compared with first
experimental data from the H1 Collaboration at HERA.
\end{quotation}
\end{titlepage}
\twocolumn
\section{Introduction}
%
The inclusive production of high-\-transverse-\-momen\-tum single hadrons
by collisions of protons and the quasi-real photons emitted by an
electron beam is an interesting process for testing the QCD-improved
parton model. First experimental results on this process have been
presented recently by the H1 Collaboration at HERA [1]. These data
have been compared already with theoretical predictions to leading
order (LO) in QCD. The LO QCD formalism consists of using the
tree-level results for the hard-scattering cross sections, the
one-loop expression for the running coupling constant, and
structure functions
and fragmentation functions generated
by one-loop evolution kernels. It is well known that such LO estimates
suffer from normalization uncertainties due to the
renormalization/factorization scale dependence. The normalization
ambiguities are substantially reduced in the next-to-leading order
(NLO) formalism.

There exist two distinct mechanisms which contribute to the inclusive
photoproduction of hadrons at high energies. The photon can interact
either directly with the partons originating from the proton, giving
rise to the ``direct'' process, or via its quark and gluon content,
the so-called ``resolved'' process. In both cases high-$\pT$ final
partons are produced, which fragment into single hadrons with the same
$\pT$.

%
Both contributions are of the same order in the perturbative
expansion. The LO contribution of the resolved part is given by LO
parton-parton scattering terms, multiplied by LO structure functions
of the photon and proton and LO fragmentation function of the final
hadron. Since the parton-parton scattering terms are of
$\O\left(\as^2\right)$ and the photon structure function
of $\O(\alpha/\as)$, both LO contributions of the direct and resolved
parts are of $\O(\alpha\as)$.

Recent data from the H1 and ZEUS Collaborations [1,2] indicate that
the resolved photoproduction dominates at HERA energies, for small
transverse momenta, $\pT$, and negative rapidities of the produced
hadron relative to the laboratory system, $\ylab$
(we take $\ylab$ to be positive in the direction of the electron beam).
So either by appropriate kinematical cuts in $\pT$ and $\ylab$ or by
observing the remnants of the resolved photon near the beam pipe it should
be feasible to obtain information on the resolved part alone [1].
Here it is understood that $\pT$ is still outside the region of soft
particle production, which dominates for $\pT\lsim1$~GeV.
Thus considering resolved photoproduction of single hadrons as a separate
process, it is important to obtain quantitative predictions for it.
For this purpose, we have calculated NLO corrections, which are of
$\O(\as^3)$ for the parton-parton scattering process, in the case of the
resolved photoproduction alone. It is clear that in the high-$\pT$ range
direct and resolved photoproduction yield comparable contributions,
which must be combined in order to obtain a reliable prediction. This
will be left for future work.

To perform a consistent calculation of inclusive single-particle
production in the NLO formalism, we need the hard-scattering cross
section in NLO, with two-loop $\as(\mu)$, two-loop evolved structure
functions of the photon and the proton, and two-loop-evolved
fragmentation functions. Only in conjunction with each other are these
elements unambiguously defined.

As far as the proton structure functions are concerned, NLO
parametrizations exist already for some time [3,4,5]. As for the photon
structure functions, recently three NLO sets of parametrizations
have appeared in the literature [6,7,8]. They have been tested
against the available experimental data on the photon structure
function ${\cal{F}}_2^\gamma(x,Q^2)$.

Unfortunately, such analyses do not exist for the fragmentation functions
of quarks and gluons in general. Only the $\pi^0$ fragmentation functions
in NLO have been developed recently [9]. This has not been done for
charged particles yet, in which we are mainly interested here.
The existing information on fragmentation functions of charged particles
is based on fits to data from $e^+e^-$ annihilation and deep-inelastic
lepton-nucleon scattering in connection with LO evolution [10,11].
These LO fragmentation functions have been tested in inclusive
charged-particle production in $p\bar p$ scattering in a previous
communication [12], henceforth referred to as {\bf I}.
Reasonably good agreement with existing data from the UA2 and CDF
Collaborations has been found.
Since the center-of-mass (CM) energies in these experiments,
$\sqrt s=(540$--1800)~GeV, are even higher than the HERA CM energy,
$\sqrt s=298$~GeV, we feel confident that these LO functions give a
reasonable account of the fragmentation of quarks and gluons into
charged hadrons.

The NLO corrections to the parton-level cross sections appropriate
for inclusive single-particle production have been worked out by
Aversa et al.\ [13].
They provide the basis of the NLO corrections to charged-particle
and neutral-pion production presented here.

In this paper we shall predict cross sections of single-charged hadron
and single-$\pi^0$ production in low-$Q^2$ $ep$ reactions
at the HERA energy.
This work is an extension of {\bf I}, where inclusive hadron
production in $p\bar p$ collisions was studied. The formalism used
here is quite similar.
One has to adjust the initial state and take care of the fact that
the incoming electron produces a spectrum of photons,
which is usually described in the Weizs\"acker-Williams
approximation (WWA) [14]. Compared to $p\bar p$ collisions, we expect in
$\gamma p$ processes a different behaviour of the cross section due
to the presence of the point-like component in the photon structure
function.

Experimental data of charged-\-particle and $\pi^0$ production at
relatively low CM energies come from fixed-target experiments at CERN
[15]. In these experiments the resolved-photon component could not be
separated from the direct one.
Resolved photoproduction cross sections for single hadrons have already
been measured at HERA [1]. A better statistic is now being accumulated
and a wider range of $\pT$ is under study.
These data will probe the QCD-improved parton model in an energy regime
that lies between $\sqrt s=200$~GeV, the $p\bar p$ CM energy of UA1,
and $\sqrt s=540$ or 630~GeV as used by UA1, UA2, and CDF.

The outline of this paper is as follows.
In Sect.~2, we give a brief introduction to the theoretical input
and the structure functions used for the calculation.
Numerical results are presented in Sect.~3, where we compare with the
first data from the H1 Collaboration.
Section~4 is reserved for a discussion of the results and some outlook
to future work.
%
\section{Theoretical ingredients and $\qquad \qquad$ structure functions}
%
When we speak of photoproduction with almost real photons we have
in mind the class of $ep$ processes where the electron is a source of
nearly massless, collinear photons, which collide with the opposing
proton beam. The energy spectrum for the resulting photons is
calculated in the WWA. In this approximation the lon\-gi\-tu\-dinal- and
transverse-momentum components of the photon distribution decouple.
The transverse momentum is integrated out with an angular cut of
$\theta_{\rm max}=5^\circ$. Then the distribution in the
longitudinal-momentum fraction, $x$, of the outgoing photon has in the
leading-logarithm approximation the following form [14]:
$$
f_{\gamma/e}(x)={\alpha\over\pi}\,{1+(1-x)^2\over x}
\ln{E_e\theta_{\rm max}\over m_e}\,,\qquad \qquad
\eqno(2.1)
$$
where $E_e$ is the electron energy, $m_e$ is the electron mass, and
$xE_e$ is the photon momentum. Equation (2.1) determines the photon
luminosity used in our calculation. When comparing with experimental
data, the actual electron acceptance must be incorporated. This will
affect our results only through a normalization factor of (2.1).
Then in the WWA the inclusive single-particle cross section is
related to the corresponding photoproduction cross section in the
following way:
$$
E_h{d^3\sigma(ep\to hX)\over d^3p_h}=\int_{x_{\rm min}}^1dx\,
f_{\gamma/e}(x)\,E_h{d^3\sigma(\gamma p\to hX)\over d^3p_h}\, \\
\eqno(2.2)
$$
On the right-hand side of (2.2), $x E_e$ is substituted for the photon
momentum and $x_{\rm min}$ is expressed in terms of transverse
momentum, $\pT$, and CM rapidity, $\ycms$, of the produced hadron as
$$
x_{\rm min} = { \pT \ e^{\ycms} \over \sqrt s- \pT \ e^{-\ycms}}\,.
\qquad \qquad \qquad \qquad \qquad
\eqno(2.3)
$$
The rapidity measured in the laboratory frame, $\ylab$, is related to
$\ycms$ by
$$
\ylab=\ycms-{1\over2}\ln{E_p\over E_e}\,,
\qquad \qquad \qquad \qquad \qquad
\eqno(2.4)
$$
where the energies of the electron and the proton, $E_e$ and $E_p$,  are
in the laboratory frame. According to current HERA conditions,
$E_e=27$~GeV and $E_p=820$~GeV, so that $\ylab=\ycms-1.71$.

In the QCD-improved parton model the single-\-par\-ticle photoproduction
cross section on the right-hand side of (2.2) is expressed in LO as a
convolution of the LO parton-parton scattering cross section with the
scale-dependent structure and fragmentation functions.
We fix the momenta of photon, proton and outgoing
hadron $(h)$ by
$$
\gamma(p_\gamma) + p(p_p) \to  h(p_h) +X\,.
\qquad \qquad \qquad \qquad
\eqno(2.5)
$$
In this way, the notation is similar to the one used in {\bf I}.
Mutatis mutandis, the LO cross section is given by
$$
E_h {d^3\sigma^0\over d^3 p_h} =
\sum_{i,j,l} \int dx_\gamma \int dx_p \int{dx_h\over x_h^2}\,
F_i^{\gamma}(x_\gamma,M^2)  \qquad
$$
$$
 \qquad{}\times
     F_j^{p}(x_p,M^2)
D_l^{h}\left(x_h,M_f^2 \right)
   k_l^0{d^3\sigma_{k_ik_j\to k_l}^0\over d^3k_l}\,,
\eqno(2.6)
$$
where $k_i = x_\gamma p_\gamma$, $k_j = x_p p_p$, and
$k_h = p_h/x_h$ are the parton momenta.
The indices $i,j,l$ run over gluons and $N_F$ flavours of quarks.
We assume $N_F=4$ throughout our calculation and include the
charm-quark
threshold as implemented in
the photon structure function parametrizations used.
$F_j^{p}(x_p,M^2)$ is the proton structure function appropriate for
parton $j$ and $D_l^{h}(x_h,M_f^2 )$ is the fragmentation function
of parton $l$ into hadron $h$.
$M$ is the factorization scale of the photon and proton structure
functions,
which we assume to be equal,
$M_f$ is the scale of the fragmentation function.
Finally, $d^3\sigma_{k_ik_j\to k_l}^0$ describes the process
$i+ j\to l+X$
in $\O\left(\as^2(\mu^2)\right)$.

The calculation of the NLO inclusive cross section is described in
{\bf I}.
The relevant formulae are used with the replacement
$F_i^{h_1}(x_1,M^2)\to F_i^\gamma(x_\gamma,M^2)$ and $\qquad$
$F_j^{h_2}(x_2,M^2)\to F_j^p(x_p,M^2)$.
For the full calculation we need the structure functions of the
photon and the proton and the fragmentation functions in NLO.
As already mentioned in the introduction, fragmentation functions
exist only in LO, except for the recently constructed $\pi^0$ set [9].
We shall use the same fragmentation functions as in {\bf I}, namely
those of Baier et al.\ [10] and Anselmino et al.\ [11] with the
constraints specified in {\bf I}. Since these fragmentation functions
correspond approximately to the \ms\ factorization, we have also
selected the \ms\ scheme for the proton and photon structure functions.

The proton structure function parametrizations are again taken from
the package by Charchu\l a [17]. For most part of our analysis we use
the Morfin-Tung (MT) set B1 in the \ms\ version.

For the NLO photon structure functions we have three sets at our
disposal [6,7,8]. All three have been tested against the available data
on the photon structure function ${\cal F}_2^\gamma(x,Q^2)$.
Unfortunately, this leaves still appreciable freedom, in particular
with respect to the gluon content of the photon. For our main results we
use the sets by Gl\"uck, Reya, and Vogt (GRV) [6] with $N_F=4$.
In their analysis the $Q^2$ evolution starts with a valence-like input
at a rather small scale, $Q_0^2$, which is $Q_0^2=0.25$~GeV$^2$
in the LO fit and $Q_0^2=0.3$~GeV$^2$ in the NLO fit, respectively.
Their NLO set refers to the so-called deep-inelastic $\gamma$ scattering
(DIS$\gamma$) scheme. For consistency in our analysis, we convert
the GRV~DIS$\gamma$ set into the more familiar \ms\ scheme.
For comparative studies at the LO level of the hard scattering we use
also the LO set by GRV. Besides, we make use also of the LO set by
Drees and Grassie (DG) [18] and the NLO set by
Aurenche et al.\ (ACFGP) [8] to investigate the sensitivity of
the cross section to different choices. The boundary conditions used
by ACFGP are similar to those of GRV and the evolution is started at
$Q_0^2=0.25$~GeV$^2$. The gluon content of the photon is particularly
low in the ACFGP set. The charm distribution is totally described by
the lowest-order expression, which has its starting point at
$Q_0^2=2$~GeV$^2$. Gordon and Storrow (GS) [7], on the other hand,
start the $Q^2$ evolution at $Q_0^2=5.3$~GeV$^2$, which is
considerably larger than the values of GRV and ACFGP.

In the next section we shall present the results in two steps.
In the first step, we assume that
$D_l^{h}(x_h,M_f^2)$ $ =\delta(1-x_h)$ for all hadrons $h$ and partons $l$.
In this way, we study the cross sections for the production of
single partons. In LO they are identical to the single-jet production
cross sections. In NLO they depend on the factorization scale
$M_f^2$ and cannot be compared to jet production cross sections, since
the recombination algorithm for defining jets is not built in.
However, the cross section with $\delta$-function fragmentation are
simpler to calculate and reveal several features that are also
relevant for the complete single-particle cross sections considered in
the second step.

\section{Numerical results}
\subsection{Single-parton inclusive cross $\qquad \qquad$ sections}
%
%
In this sub-section we approximate all fragmentation functions by $\delta$
functions. This leads to single-parton inclusive cross sections.
For a first orientation, we disentangle the various parton-parton
channels that contribute in LO to the hard-scattering cross sections
ordering them according to their initial state, viz.\
\begin{itemize}
\item[{1)}] $gg\to gg,\ q\bar q$,
\item[{2)}] $gq\to gq,\ qg$,
\item[{3)}] $qg\to gq,\ qg$,
\item[{4)}] $q\bar q\to gg,\ q\bar q,\ \bar qq,\ q^\prime\bar q^\prime$,
\hfil\break
$qq\to qq$,
\hfil\break
$qq^\prime\to qq^\prime,q^\prime q$,
\end{itemize}
\noindent
where $q$ and $q^\prime$ denote different quark flavours.
The first parton originates from the photon, the second from the
proton. These four classes of channels are also compared with direct
photoproduction in LO,which proceeds through $\gamma g\to q\bar q$
and $\gamma q\to gq,\ qg$. The comparison is done for
$d^2\sigma/dy\,d \pT$ at $\sqrt s=298$~GeV (this corresponds
to 27~GeV electrons colliding with 820~GeV protons according to
present HERA conditions) and $\ylab=-2$ (with $\ylab>0$ in the
direction of the electron momentum). For this calculation we use the LO
set SL of the MT
proton structure functions and the LO set of the GRV photon
structure functions. We also fix the two scales $\mu$ and $M$ to
be $\mu=M=\pT$.

In Fig.~1 we show the $\pT$ dependence of the channels 1)-4) contributing
to the resolved $\gamma p$ cross section and the $\pT$ dependence of the
direct process. We observe that at small $\pT$ channels 1),~2), and 3)
dominate by one order of magnitude over class 4) and the direct
mechanism, whereas at larger $\pT$ ($\pT \gsim 28$~GeV) class 4), in
particular the channels involving quarks only, contribute significantly.
It is exactly in this $\pT$ range, more precisely for $\pT > 35\,$GeV,
that the direct process exceeds the sum of the resolved
processes. The pattern observed for the various resolved photoproduction
channels is similar to the situation encountered in {\bf I} in the case
of $p\bar p$ collisions: the gluon components of the photon and proton
structure functions influence strongly the cross section at smaller
$\pT$, whereas the pure quark channels dominate at larger $\pT$.

In Fig.~2, the rapidity distributions of the above
channels are displayed for fixed $\pT=5$~GeV.
For $\ylab<0$, i.e., in the proton direction, classes 1) and 3)
dominate and the direct mechanism comes in only in the extreme forward
direction, at $\ylab\approx 1$.
Of course, the relative size of the various classes in Figs.~1,2 may
change when realistic fragmentation functions are taken into account.

\begin{figure}[tb]
\epsfxsize=8.0cm
\epsfbox[20 74 556 731]{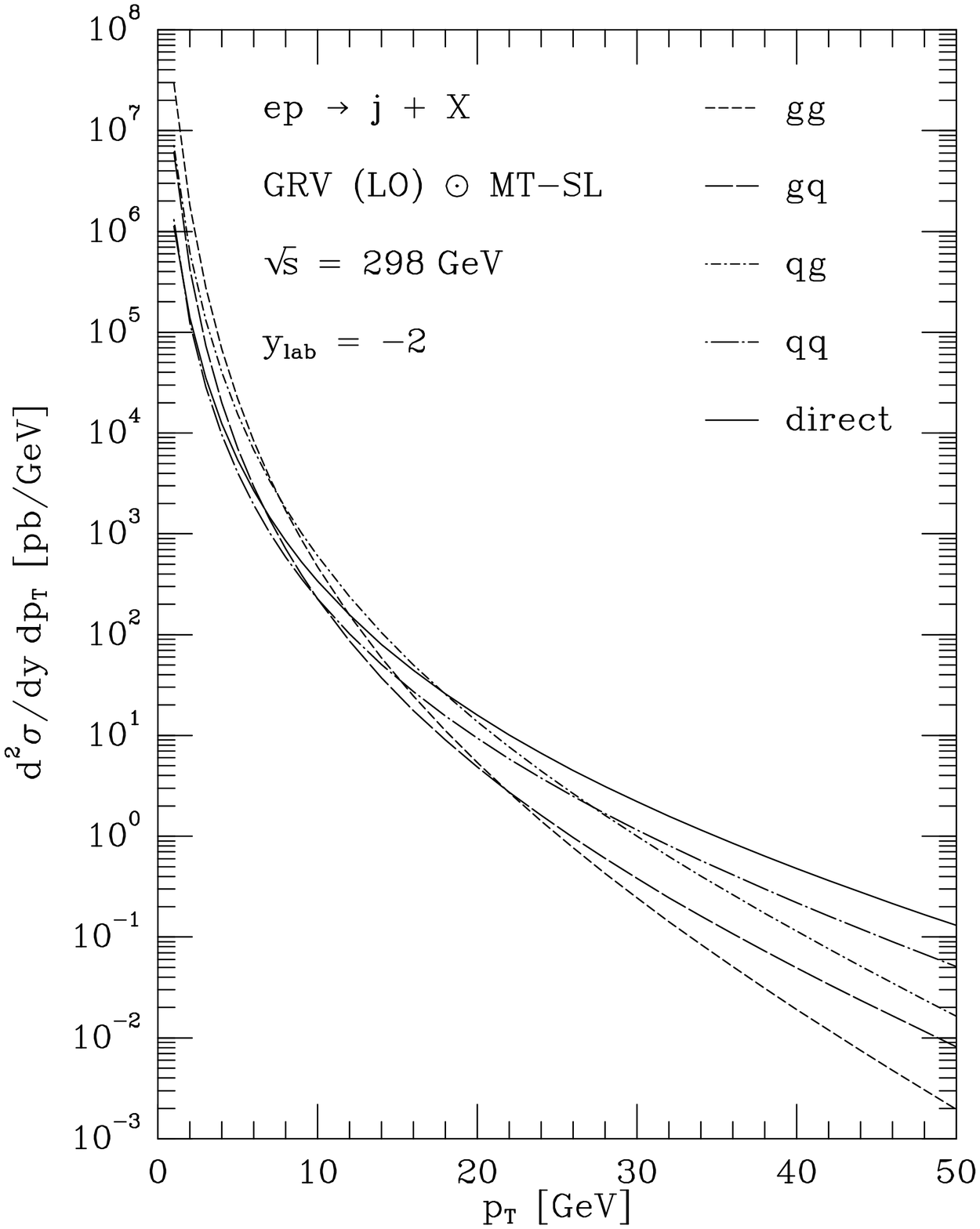}
{{\small {\bf Fig.~1.} Transverse-momentum distributions of the
various channels for
resolved photoproduction and of direct photoproduction in LO at
$\sqrt s=298$~GeV and $\ylab=-2$ evaluated with the LO sets by GRV
and MT-SL. The resolved-photon channels are grouped according to their
parton-parton initial state; the first parton originates from the photon
and the second from the proton}}
\end{figure}

\begin{figure}[tb]
\epsfxsize=8.0cm
\epsfbox[20 74 556 731]{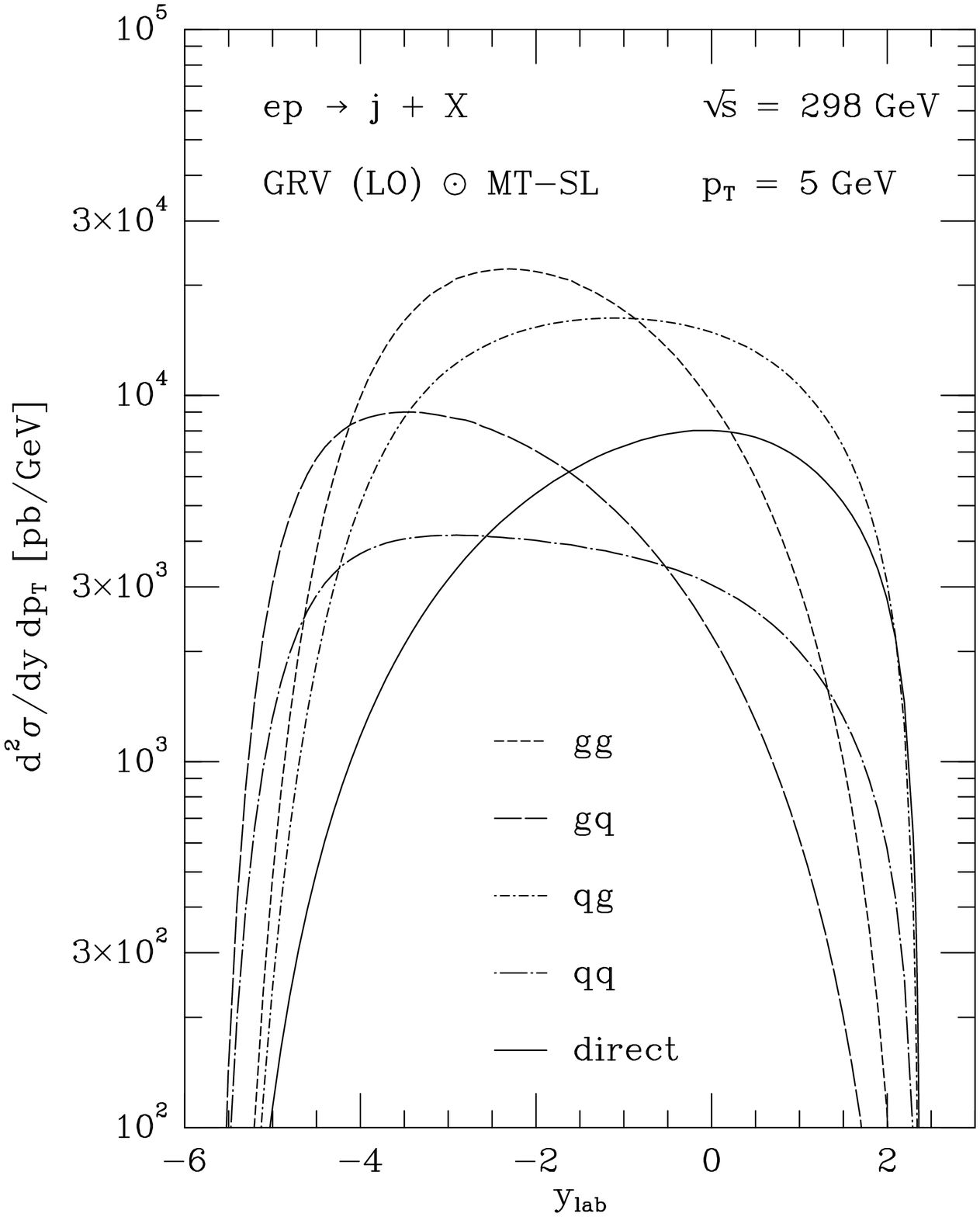}
{{\small {\bf Fig.~2.} Rapidity distributions of the various
channels for resolved
photoproduction and of direct photoproduction in LO at $\sqrt s=298$~GeV
and $\pT=5$~GeV evaluated with the LO sets by GRV and MT;
see caption of Fig.~1}}
\end{figure}

Next we study how our results are changed when different sets of proton
structure functions are employed. Toward this end, we calculate
$d^2\sigma/dy\,d\pT$ in the Born approximation at $\sqrt s=298$~GeV
and $\ylab=-2$ as a function of $\pT$ using in turn the
proton structure function sets MT--B1~(DIS), MT--B1~(\ms) [3] and
MRSn--D0~(\ms), MRSn--D$-$~(\ms) [5], which are all of NLO.
For the photon structure functions we choose the
LO set by GRV [6] and for the QCD coupling
$\as$ we use the usual one-loop formula adopting the value
of $\Lambda$ from the respective proton structure functions.
In Fig.~3, we show the cross sections normalized with respect to the
calculation with the MT--SL set.
We remark that only the MT--SL set should be combined with the LO
hard-scattering cross sections because it is a genuine LO fit;
all the others correspond to NLO analyses of present data.

\begin{figure}[tb]
\epsfxsize=8.0cm
\epsfbox[51 320 564 729]{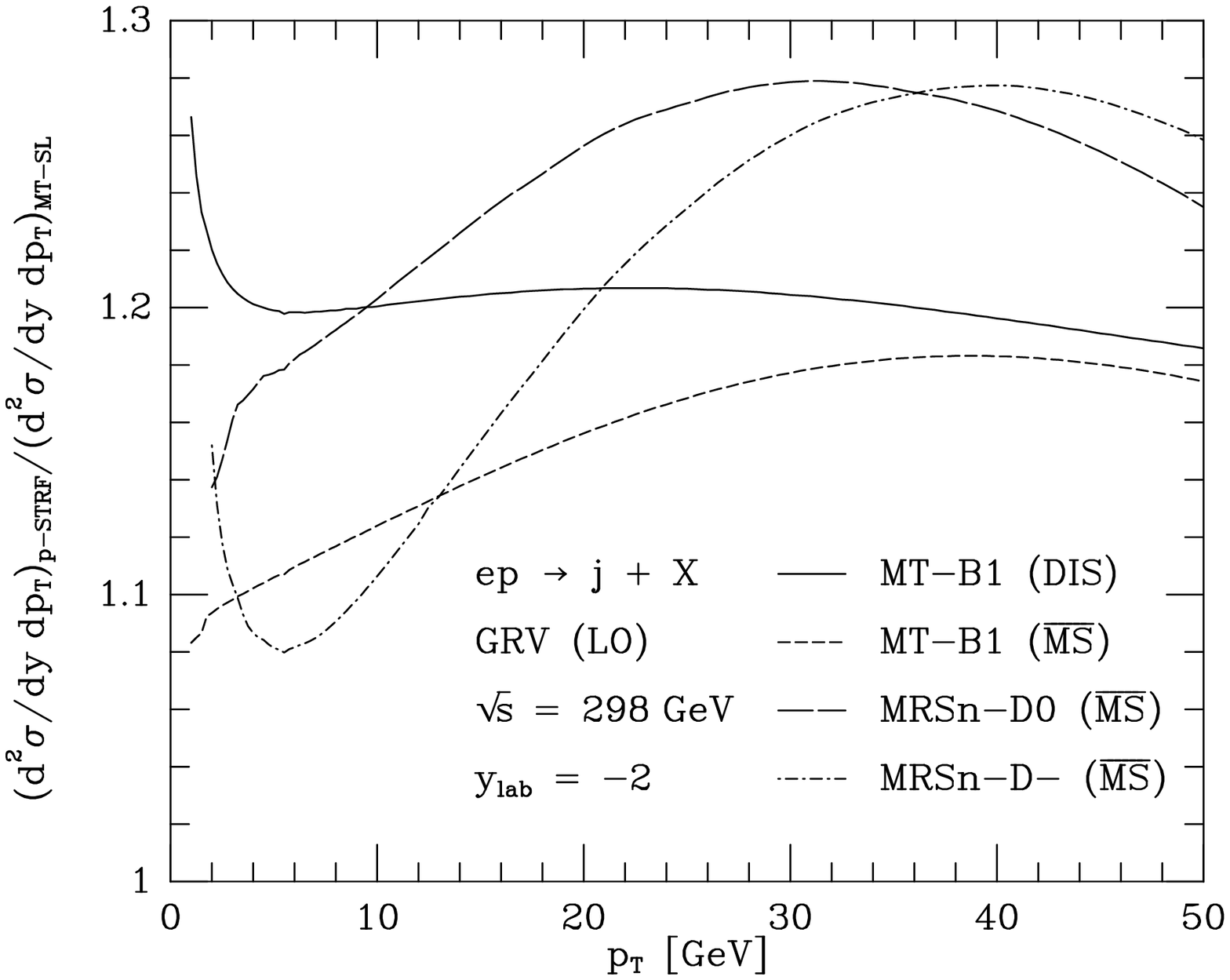}
{{\small {\bf Fig.~3.} Influence of typical NLO proton structure
functions on the
LO calculation of $d^2\sigma/dy\,d \pT$ for resoved photoproduction
at $\sqrt s=298$~GeV and $\ylab=-2$.
The curves are normalized to the LO calculation with the MT--SL set.
In both numerator and denominator the GRV LO photon structure functions
are used}}
\end{figure}

We observe that the variation caused by replacing the LO set by the
various NLO sets does not exceed 30\% for $\pT\gsim2$~GeV. A similar
result was found in {\bf I} for $p\bar p$ processes. We observe that
the ratio of the two cross sections is always larger than one. From
what was said above, it follows that the comparison of the NLO sets
with the LO set is not particularly meaningful.
However, it is instructive to compare the results obtained with
different NLO sets of proton structure functions parametrizations.
The corresponding cross sections differ at most by 20\%.
The DIS and \ms\ versions of MT--B1 differ by only 10\% relatively to
each other. We consider this as an upper bound on the scheme dependence,
since the partial compensation that is expected to take place at NLO,
due to scheme-dependent terms, is not in effect here.
The MRSn--D0 and MRSn--D$-$ sets fit also recent NMC [19]
and CCFR [20] data, which extend to smaller values of $x$ than
previously tested. The D0 and D$-$ sets, available in
the \ms\ version, give cross sections
with quite different behaviours at low $\pT$. At large $\pT$ they give
similar results and differ from the cross section obtained making use of
the MT--B1~(\ms) set by (10--15)\%. When comparing with the pattern
observed for $p\bar p$ reactions (see Fig.~2 in {\bf I}), we must keep
in mind that there the structure functions of both proton and antiproton
are changed, whereas here the photon structure function is left unchanged.

To study the influence of different photon structure functions, we
calculate $d^2\sigma/dy\,d \pT$ again in the Born approximation at
$\sqrt s=298$~GeV and $\ylab=-2$ as a function of $\pT$ using in turn
the NLO sets GRV~(DIS$\gamma$), GRV~(\ms) [6], ACFGP (massive) [8] and
GS [7] and the LO parametrizations GRV [6], DG [18] and GS [7]. In all
cases the MT--SL proton structure function parametrization is used. The
results are shown in Fig.~4.
Except for low $\pT$, the DG set gives cross sections which are 10\% smaller
than those for GRV (LO).
The two steps in the DG curve are due to the fact that the
DG parametrization is discontinuous at the charm and bottom thresholds,
at $\pT\approx5$~GeV and $\pT\approx14$~GeV, respectively.
The GRV structure functions are smooth at these thresholds.
The GRV~(DIS$\gamma$) set gives results very close to the GRV (LO) results.
This set involves a particular subtraction in the point-like part of the
photon structure function, which is compensated by a change of the direct
photoproduction cross section. Since the latter is beyond the scope of
this work, we shall not employ the GRV~(DIS$\gamma$) set in the following.
We observe that the DIS$\gamma$ and \ms\ versions of the GRV parametrization
differ appreciably
only at larger $\pT$, where direct photoproduction comes into play;
at small $\pT$ ($\pT\lsim25$~GeV) their difference is below 10\%.
The cross section for the ACFGP choice of structure functions,
which is also based on the \ms\ scheme, is very different, especially in the
lower-$\pT$ range, where it is up to 50\% smaller than the GRV results.
This can be traced to the much weaker gluon component in the ACFGP set.
At large $\pT$, where the influence of the gluon component diminishes,
the ratio approaches the value 1 again.
The results obtained for the LO and NLO GS photon structure function
are not included in Fig.~4. The LO soft-gluon option yields values for
the ratio between 1 and 1.2 for $\pT\lsim16$~GeV.
In the upper range of the $\pT$ values considered in Fig.~4 the ratio comes
down to 0.9. The NLO set, which is based on the \ms\ scheme, gives
results in the ballpark of those obtained with the DG and ACFGP sets;
for $\pT\lsim20$~GeV the GS~(\ms) values are somewhat smaller than the DG
values,
for $\pT\gsim30$~GeV they are in line with the ACFGP values.

\begin{figure}[tb]
\epsfxsize=8.0cm
\epsfbox[51 320 564 729]{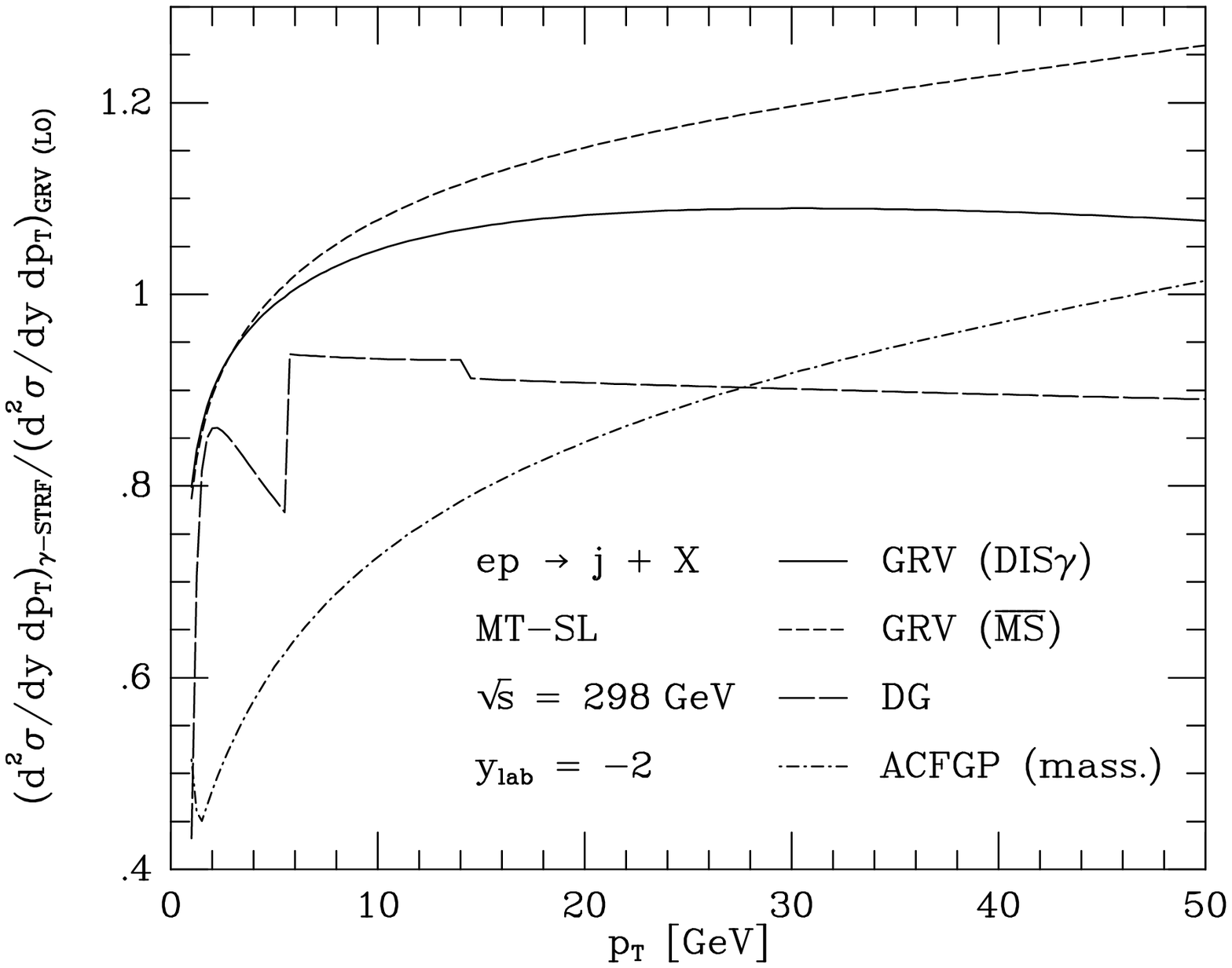}
{{\small {\bf Fig.~4.} Influence of typical photon structure functions on the
LO calculation of $d^2\sigma/dy\,d \pT$ for resoved photoproduction
at $\sqrt s=298$~GeV and $\ylab=-2$.
The curves are normalized to the LO calculation with the GRV LO set.
In both numerator and denominator the MT--SL proton structure functions
are used}}
\end{figure}

To investigate the influence of the higher-\-order $\qquad$ terms of the
hard-scattering
cross sections, we restrict ourselves to pure photoproduction,
replacing the right-hand side of (2.1) by $\delta(1-x)$.
We calculate $d^2\sigma(\gamma p \to h X)/dy\,d \pT$
at $\sqrt s=256$~GeV (corresponding to typical 20~GeV photons colliding
with 820~GeV protons) and $\ylab=-2$ as a function of $\pT$ with and
without the higher order terms and plot the ratio in Fig.~5.
This ratio should change only little when the photon spectrum (2.1) is
taken into account.
A realistic analysis including the WWA and also fragmentation will be
reported
in the next sub-section.
At this point, we wish to separate the effect of the higher-order terms.
To that end, we use the same structure functions, namely the \ms\ versions
of GRV and MT--B1, and the two-loop expression for $\as$ in both
numerator and denominator.
The renormalization scale of $\as$ is adjusted according to the proton
structure functions, i.e., $\Lambda=194$~MeV, which is very close to
the value $\Lambda=200$~GeV ($N_F=4$) implemented in the GRV set.
We perform the computation for the three choices $\mu=M=M_f=\pT/2$, $\pT$,
and $2 \pT$. From
Fig.~5 we see that the ratio is practically independent of $\pT$
for $\pT\gsim10$~GeV.
For smaller $\pT$ it increases with decreasing $\pT$ at a rate that is
largest for the smallest scale.
Its value ranges between 1.5 and 2.5, which demonstrates the importance
of the NLO corrections to the parton-level cross sections.
These results are very similar to the ones reported in {\bf I}
for $p\bar p$ reactions at $\sqrt s=630$~GeV.
Note that, in contrast to Figs.~3 and 4, the same structure
functions are used
in both numerator and denominator, so that variations due to changes of
scheme or parametrization largely cancel out.
We observe a strong scale dependence, which is mainly due to the
denominator, as will become apparent below.

\begin{figure}[tb]
\epsfxsize=8.0cm
\epsfbox[51 320 564 729]{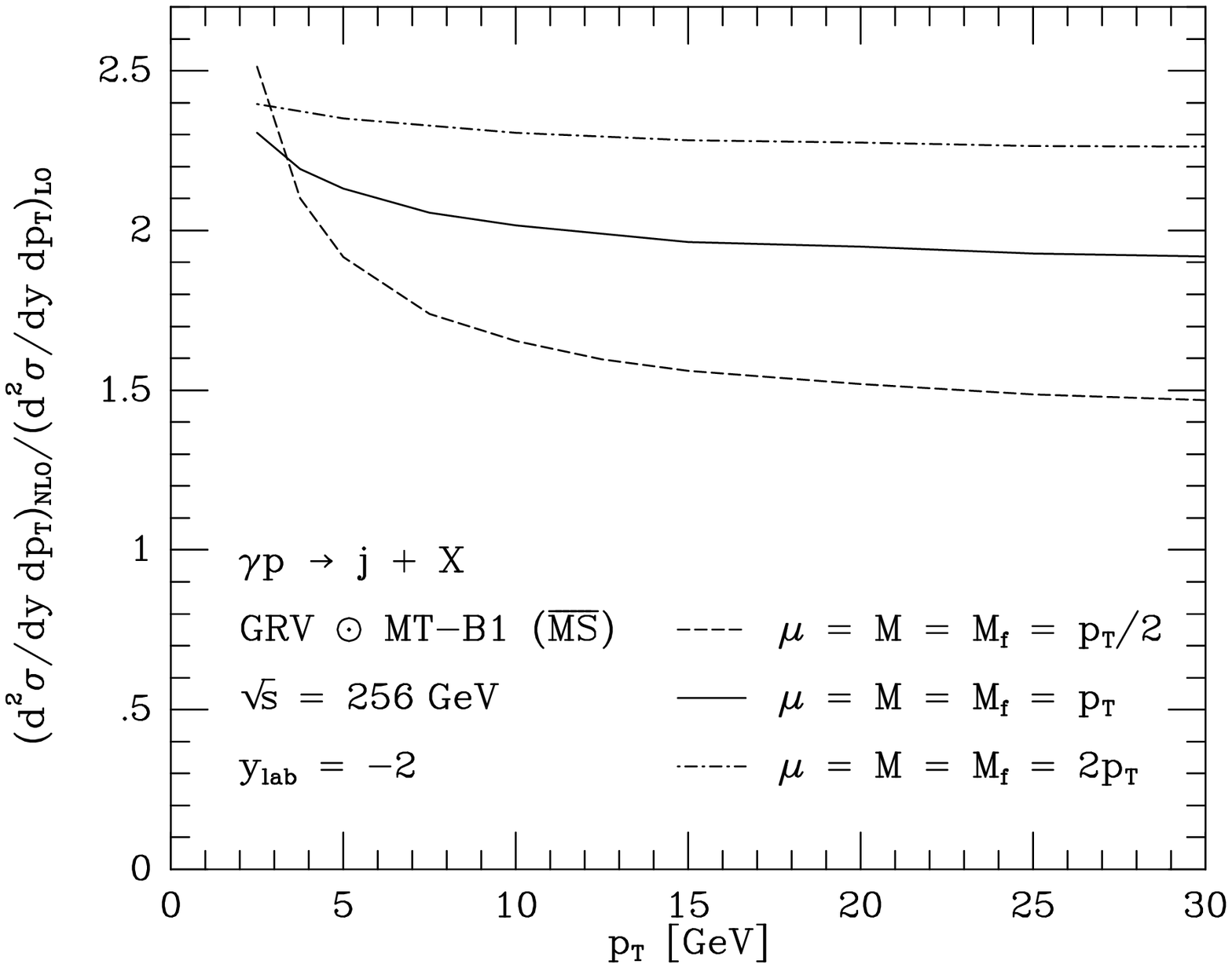}
{{\small {\bf Fig.~5.}
NLO calculation in the \ms\ scheme of
$d^2\sigma/dy\,d \pT$
for $\gamma p\to j+X$ at $\sqrt s=256$~GeV and $\ylab=-2$ normalized
to the corresponding calculation with the NLO terms in the hard-scattering
cross section omitted.
In both numerator and denominator the \ms\ sets of the GRV and MT--B1
structure functions are used.
The scales $\mu$, $M$, and $M_f$ are identified and set equal to
$\pT/2$, $\pT$, and $2\pT$}}
\end{figure}

At this point, a few comments on spurious higher-order effects are in order.
In Fig.~5, we used NLO structure functions also in connection with the
NLO corrections to the hard-scattering cross sections.
Strictly speaking, such a procedure introduces certain contributions
beyond next-to-leading order in the case of GRV.
For the choice $\mu=M=M_f=\pT$ we have explicitly checked the influence of
the modifications necessary to eliminate these spurious higher-order terms.
At small $\pT$ ($\pT\lsim5$~GeV), the ratio is increased by some 10\%,
but already at $\pT \gsim 7\,$GeV the increase becomes insignificant, until
it vanishes at $\pT \sim 12\,$GeV. Then the ratio is reduced by an
amount which reaches at most 5\% at the upper end of the $\pT$-range
shown in Fig.~5. Similar results are obtained for single-hadron
production: the only effect of the realistic fragmentation functions is
to reduce the values of $\pT$ at which the features previously described
take place.
Since at present there is still appreciable freedom concerning the details
of the photon structure functions, we do not consider it necessary to
worry about
this complication,
leaving it for a future, more refined analysis.

\begin{figure}[tb]
\epsfxsize=8.0cm
\epsfbox[51 320 564 729]{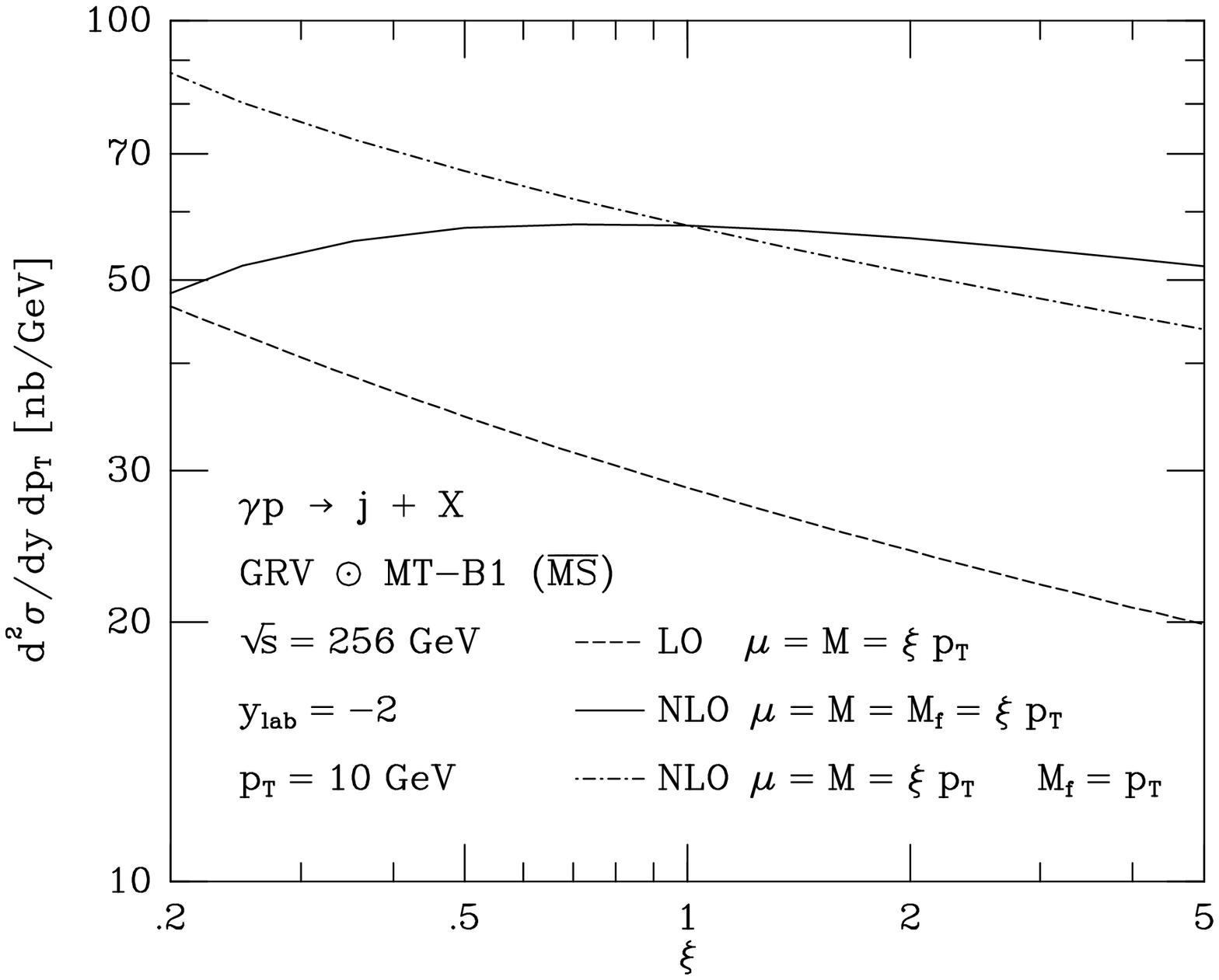}
{{\small {\bf Fig.~6a.} Scale dependence of
$d^2\sigma/dy\,d \pT$ for $\gamma p\to j+X$
at $\sqrt s=256$~GeV, $\ylab=-2$, and $\pT=10$~GeV.
The solid and dot-dashed lines represent the NLO results in the \ms\ scheme
with variable and fixed $M_f$, respectively.
The \ms\ sets of the GRV and MT--B1 structure functions are used.
The dashed lines correspond to the calculation with the NLO terms of the
hard-scattering cross sections omitted}}
\end{figure}
\setcounter{figure}{5}

\begin{figure}[tb]
\epsfxsize=8.0cm
\epsfbox[51 320 564 729]{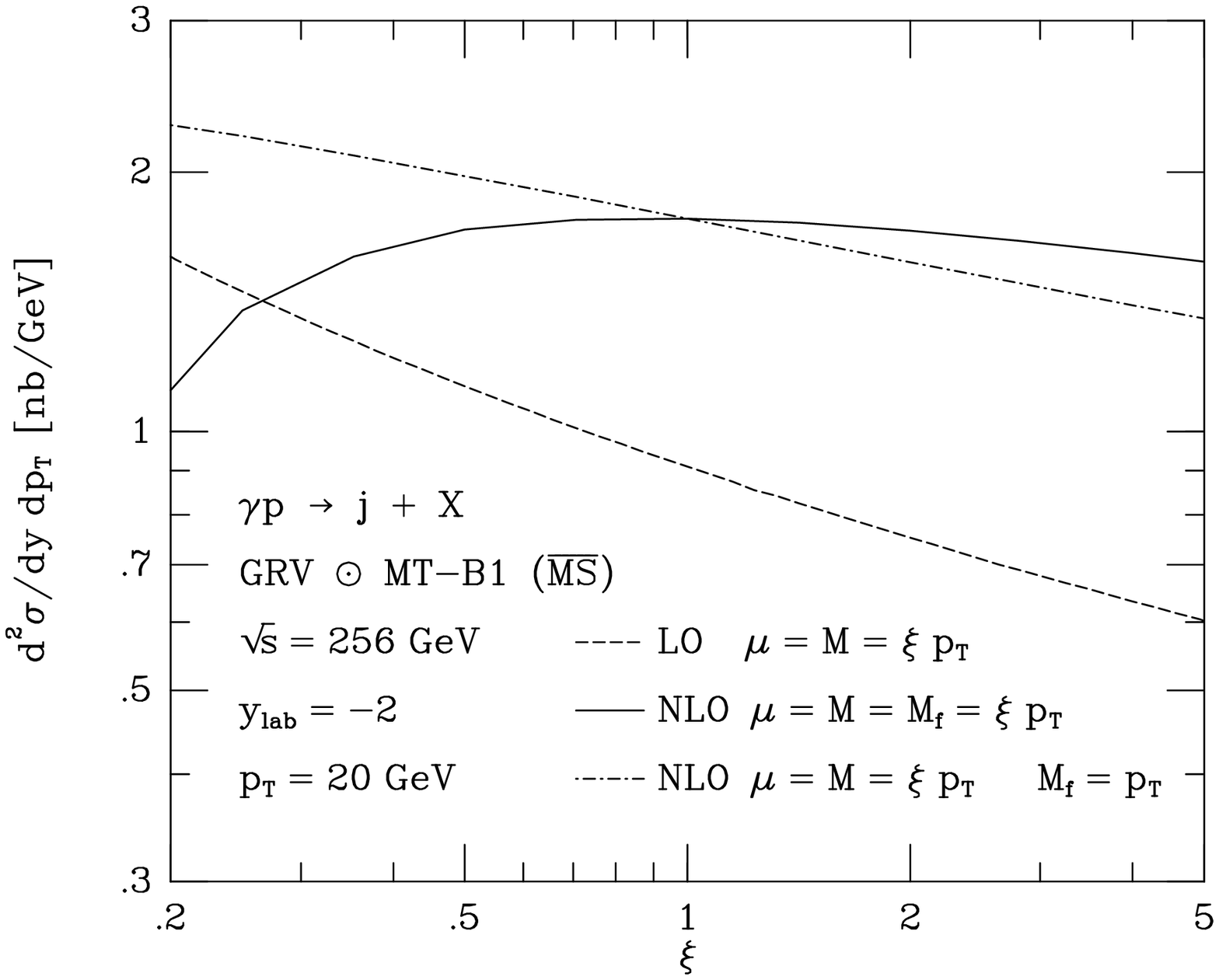}
{{\small {\bf Fig.~6b.} Same as in Fig.~6a but at $\pT=20$~GeV.}}
\end{figure}

The scale dependence of numerator and denominator of the ratio plotted
in Fig.~5 is explicitly investigated in Figs.~6a,b for $\pT$ fixed at
10 and 20~GeV, respectively.
The three scales $\mu$, $M$, and $M_f$ are set equal to $\xi \pT$ and
$\xi$ is varied between 0.2 and 5.0.
Again, we choose the \ms\ sets by GRV and MT, $\sqrt s=256$~GeV, and
$\ylab=-2$.
The LO cross section (dashed line) depends more strongly on the scale
$\xi$ than the NLO result (solid line).
The more rapid decrease of the LO cross section for increasing $\xi$
explains the increase of the ratio plotted in Fig.~5.
The NLO cross section is particularly insensitive to $\xi$ for $\xi>0.5$.
This holds true independently of $\pT$, although the absolute cross
section strongly varies with $\pT$.
The scale variation of the NLO cross section is well bounded in contrast
to the LO cross section, which decreases monotonically with increasing
$\xi$.
At this point, one should keep in mind that the $M_f$ dependence in the
NLO hard-scattering cross section is not compensated as long as
$\delta$-function-type fragmentation is assumed.
If we do not vary $M_f$, which would be more appropriate in this case,
we obtain the dot-dashed curves in Figs.~6a,b. This shows less $\xi$
dependence than the LO curves but more than the NLO curves with
variable $M_f$.
It seems that the factorization scale dependence in the NLO hard-scattering
cross section is mostly cancelled between initial and final state.
The pattern observed in Figs.~6a,b is very similar to our findings
regarding $p\bar p$ collisions in {\bf I}.

\subsection{Single-hadron in\-clu\-sive cross $\qquad \qquad$ sections}
%
In this sub-section we shall present our results for the inclusive
pro\-duc\-tion of sin\-gle charged hadrons and neutral pions. For the
calculation of these cross sections we used the same
frag\-men\-ta\-tion func\-tions as in Sect.~2. We choose the MT-B1
and GRV structure functions, both in the \ms\ scheme, except in
Fig.~7, where we take the MT-SL and GRV~(LO) structure functions.
As in the previous sub-section, the parameter
$\Lambda$ in the coupling constant $\as$ is set equal to the value
required by the proton structure function parametrizations, i.e.
194\,MeV for NLO evaluations and 144\,MeV for the LO ones.

\begin{figure}[tb]
\epsfxsize=8.0cm
\epsfbox{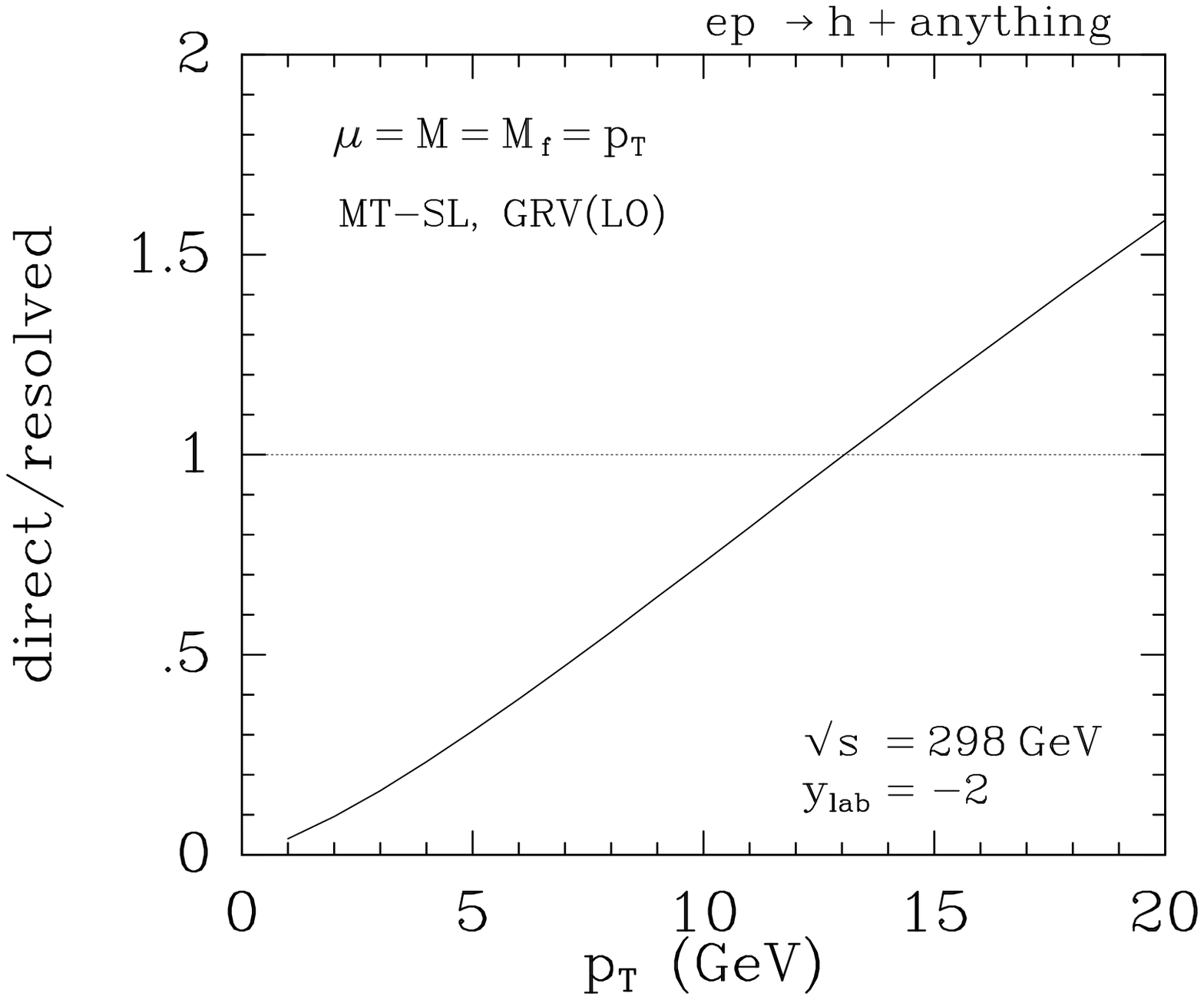}
{{\small {\bf Fig.~7.} Ratio of direct over resolved contribution
to the cross section for production of charged hadrons  }}
\end{figure}

First we consider the ratio of direct over resolved production of
charged hadrons $h\equiv (h^+ + h^-)/2$ in order to see where they
are equal in magnitude. We restrict ourselves to the LO
calculation and we set $\sqrt{s}=298\,$GeV, $\ylab =-2$, and
$\mu=M=M_f=\pT$. As in the previous sections, $M$ indicates both,
the factorization scales for the photon and the proton structure
functions. The result is shown in Fig.~7. The crossing point, above
which the direct production of single hadrons exceeds the resolved
one, is at $\pT \sim 13\,$GeV. This value is considerably lower than
the crossing point
obtained in
the case of LO jet-production cross section.
We do not expect the position of these points to change drastically
when NLO corrections are included. Fig.~7 shows clearly that
for $\pT>5\,$GeV the direct production of single charged hadrons is
not negligible anymore and our results can only apply to data with a
visible photon remnant jet.

\begin{figure}[tb]
\epsfxsize=8.0cm
\epsfbox{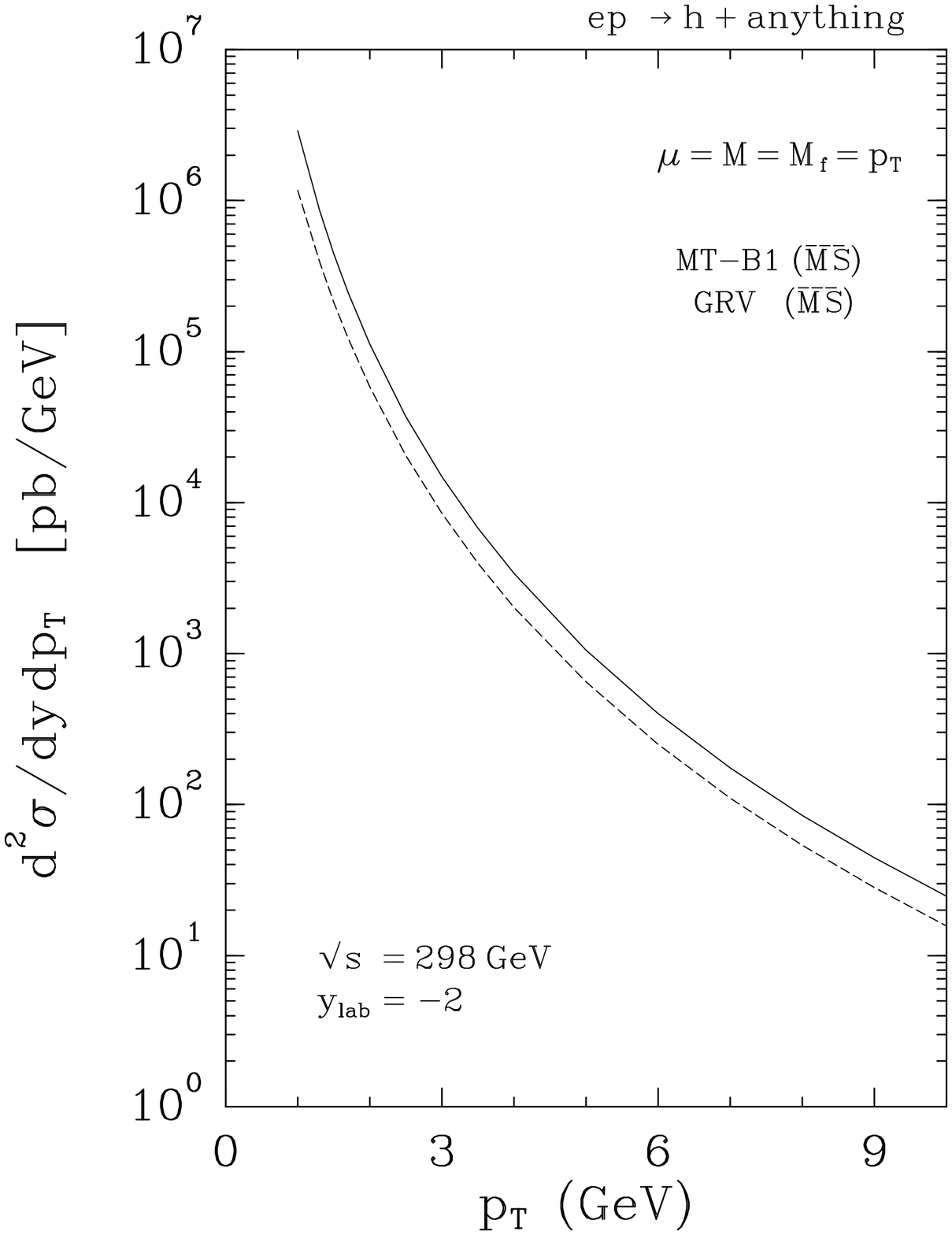}
{{\small {\bf Fig.~8.} Inclusive charged-hadron production cross section
for $\sqrt{s}=298\,$GeV and $\ylab=-2$. The solid line correspond
to the NLO prediction while the dashed line is the LO result obtained
with structure functions and $\as$ kept at the NLO level } }
\end{figure}

\begin{figure}[tb]
\epsfxsize=8.0cm
\epsfbox{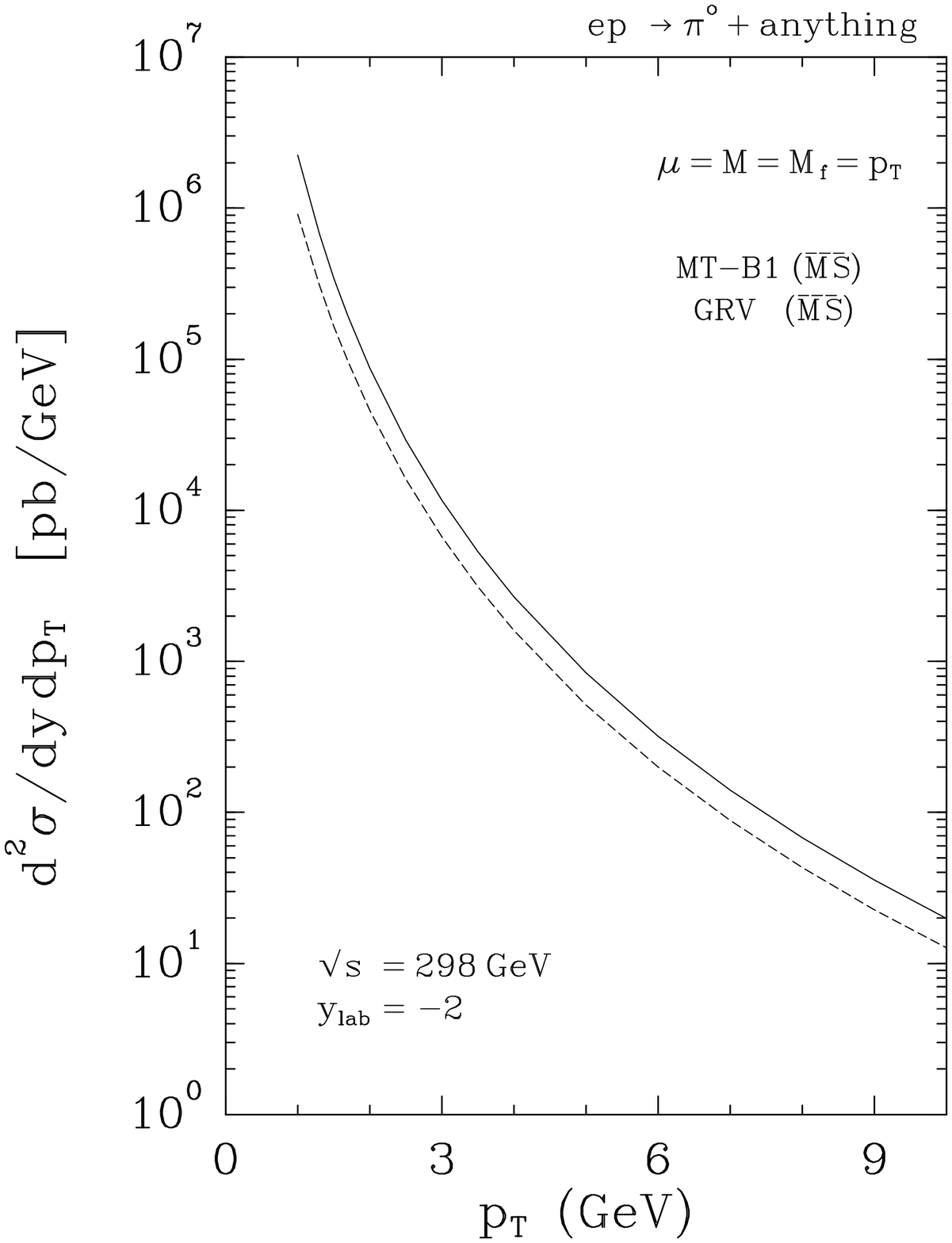}
{{\small {\bf Fig.~9.} Same as in Fig.~8 for single-$\pi^0$ production }}
\end{figure}

In Fig.~8 we show $d^2\sigma/dy\,d\pT$ for inclusive
single-charged hadron production as a function of
$\pT$ for $\sqrt{s}=298\,$GeV, $\ylab=-2$, and $\mu=M=M_f=\pT$.
The upper curve is the NLO result and the lower curve the LO cross
section with the same structure functions, MT-B1 and GRV in
$\overline{\rm MS}$, and $\as(\mu^2)$. K factors ranging between
2.5 and 1.6 are obtained in the $\pT$ range considered here. They are
smaller at hight $\pT$ than the factors observed
in the case of $\delta$-function fragmentation (see Fig.~5),
for the same choice of scales,
but bigger at high $\pT$. Moreover, the inclusion of realistic
fragmentation functions for the produced hadrons induces a decrease of
the cross section, for increasing $\pT$, faster than the one observed
in Fig.~1.

Qualitatively similar results, shown in Fig.~9, are obtained in the
case of single-$\pi^0$ production. However, the values of the cross
sections are here lower than the ones obtained in the case of
charged-hadron production, for the same choice of scales. It should
be reminded at this point that the results presented in {\bf I} for
single-$\pi^0$ production in $p\bar{p}$ collisions could fit the
existing data only for relatively high scales. Similar conclusions
were reached in [9] when the $\pi^0$-fragmentation functions [11]
were used.

\begin{figure}[tb]
\epsfxsize=8.0cm
\epsfbox{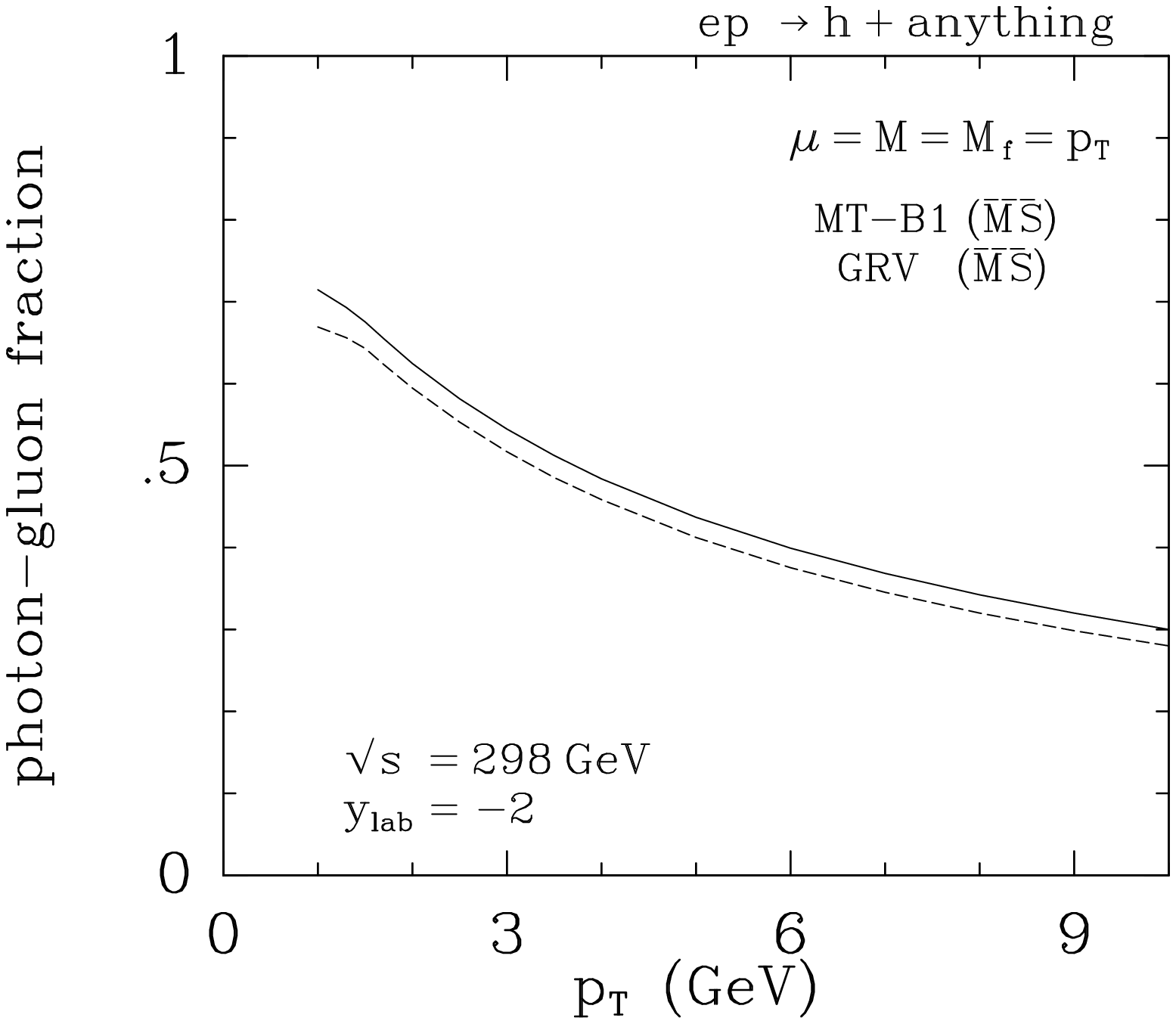}
{{\small {\bf Fig.~10.} Fraction $R_{g/\gamma}$
of the charged-hadron production cross section at
$\sqrt{s}=298\,$GeV and $\ylab=-2$ due to the gluon originating from the
photon. The solid line indicates the NLO
result and the dashed line the LO one with structure functions
and $\as$ at the NLO level} }
\end{figure}

In connection with the results in Figs.~8 and~9, it is of interest to
know the influence on the cross section of the gluon originating from
the photon. According to the results shown in Fig.~1, we expect this
gluon contribution to be more important at small $\pT$. In
Fig.~10 we plot the $\pT$ dependence of the ratio
$$
R_{g /\gamma} \equiv
{d^2\sigma/dy\,d\pT - d^2\sigma/dy\,d\pT\vert_{({\rm no}\ g/\gamma)}
 \over
 d^2\sigma/dy\,d\pT}\,, \qquad
\eqno(2.7)
$$
where the label ${({\rm no}\ g/\gamma)}$ indicates that the gluon
component in the photon has been put to zero. $R_{g/\gamma}$ varies from
$60 \%$ at $\pT=2\,$GeV to $30\%$ at $\pT=10\,$GeV. This result
is essentially independent of whether NLO corrections are included or not.

\begin{figure}[tb]
\epsfxsize=8.0cm
\epsfbox{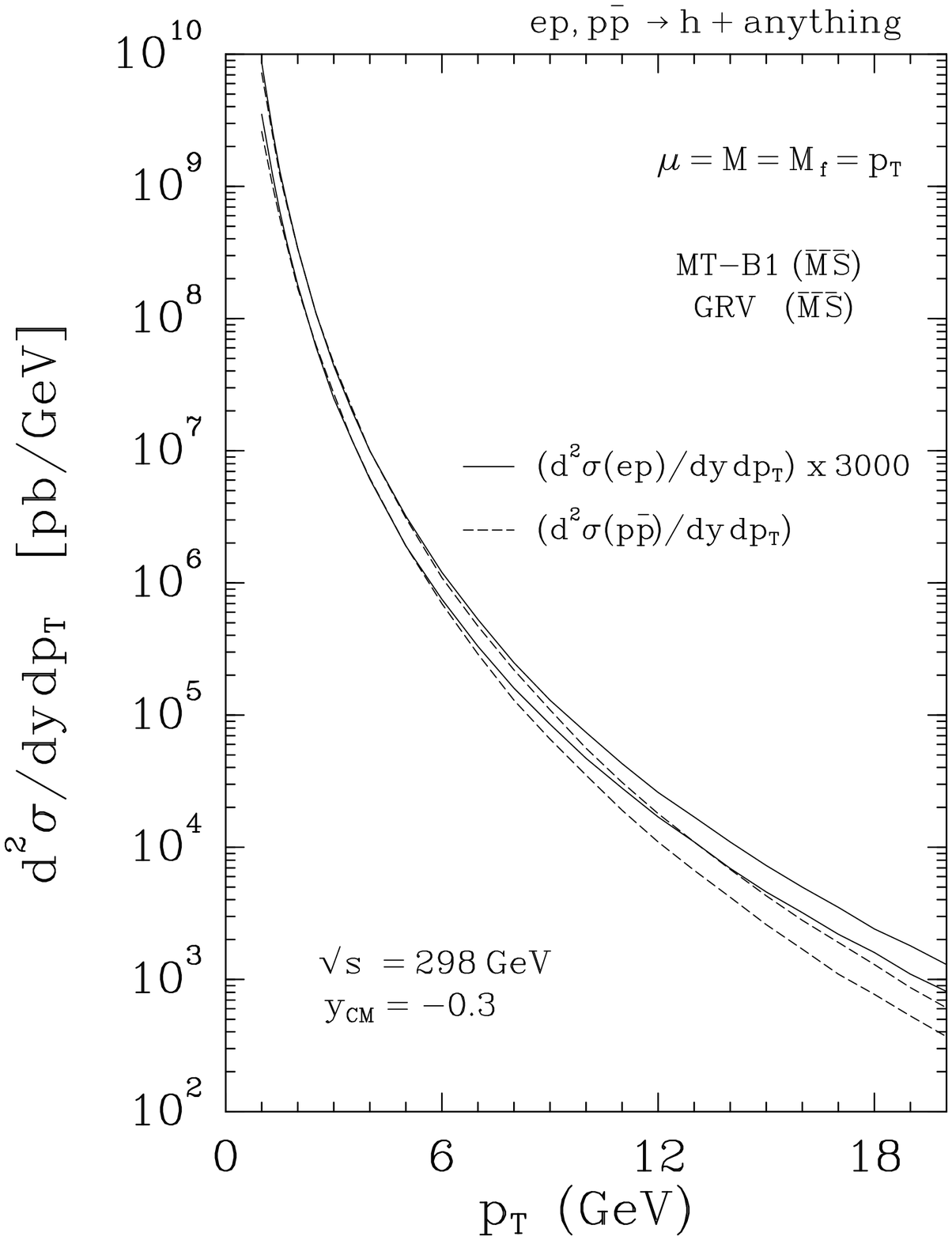}
{{\small {\bf Fig.~11.} Comparison of the
cross sections of charged-hadron production
in $p\bar{p}$ collisions and via resolved photons in $ep$ collisions
at HERA energy and the same CM energy and rapidity, $\sqrt{s}=298\,$GeV
and $\ycms=-0.3$. The upper curves indicate the full NLO results, while
the lower ones give the LO results obtained with structure function and
$\as$ at the NLO level}}
\end{figure}

The importance of the quark component of the photon for increasing
$\pT$ should become evident when we compare single-charged hadron
production in $ep$ and $p \bar{p}$ collisions at the same
$\sqrt{s}$. Since, at large $\pT$, the point-like part of the
photon-quark
component is much stronger than the
proton-quark
distribution, we expect the $p\bar p$ cross section to fall off more
rapidly than the $ep$ cross section, for increasing $\pT$. This is
indeed the case, as can be seen in Fig.~11 where we compare
$ ep \to h X$ via resolved photons with $p \bar{p} \to h X$ at
$\sqrt{s}=298\,$GeV and a CM rapidity at which both cross sections
are maximal, $\ycms=-0.3$. The two curves in Fig.~11 are NLO
(upper curves) and LO (lower curves) predictions, respectively. The
values of the $ep$ cross section have been multiplied by the factor
3000 which makes $ep$ and $p \bar{p}$ results coincide up to
$\pT\sim 4\,$GeV. Already at $\pT= 10\,$GeV, though, the two sets of
curves deviate visibly from each other and at the upper end of the
$\pT$-range shown in Fig.~11 the $p \bar{p}$ cross section falls
short of the $ep$ one by at least a factor two. Of course, in order
to verify this effect experimentally, we need $p \bar{p}$ and
$ep$ data at the same $\sqrt{s}$ and the isolation of the resolved
production in the $ep$ process. With direct $ep$ production
included, the effect should be even stronger.

Another issue to be considered before proceeding further is whether
the inclusion of a fifth flavour introduces any appreciable
deviation from
the results presented so far. We have verified that, in spite of the
non-negligible contribution of the bottom-quark to the photon structure
function, the increase in the cross sections is at most 2\%, for the
highest $\pT$ values here considered, when we assume $N_f=5$.

\begin{figure}[tb]
\epsfxsize=8.0cm
\epsfbox{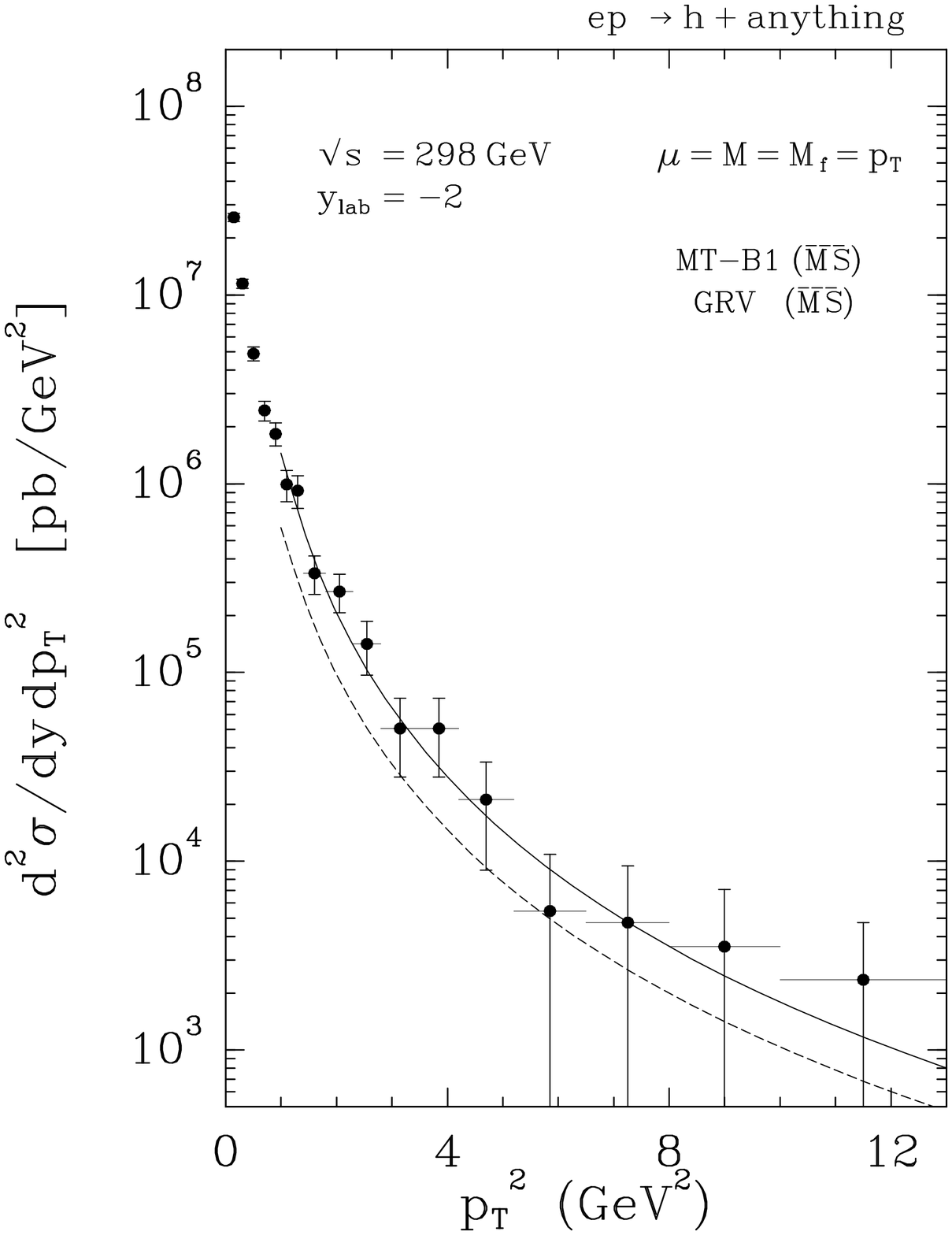}
{{\small {\bf Fig.~12.} Comparison between the theoretical predictions for
the resolved-photon contribution to $ep \to hX$ and the experimental
data obtained by the H1 Collaboration at HERA [1]. The data are normalized
globally in such a way that they match with the NLO prediction at
$\pT^2\sim1.4$--$1.7\,$GeV$^2$. The solid line indicates the NLO
prediction and the dashed line the LO one obtained with
structure functions and $\as$ at the NLO level }  }
\end{figure}

Finally, in Fig.~12 we com\-pare our pre\-dic\-tions from Fig.~8 with
recent data obtained by the H1 Collaboration at HERA [1]. These data
extend only up to $\pT^2=12\,$GeV$^2$ and have quite large
ex\-per\-i\-men\-tal un\-cer\-tain\-ties for the largest values of
$\pT^2$. Since the absolute normalization of these data points is not
known, we have adjusted them with an overall factor so as to have
satisfactory agreement with the theoretical NLO results at
$\pT^2\sim 1.4$--$1.7\,$GeV$^2$. We see that the $\pT^2$ shape of
the the\-o\-ret\-i\-cal re\-sults is con\-sis\-tent with the
data. Of course, for firmer conclusions, more accurate data which
extend to higher values of $\pT^2$ are needed.

\begin{figure}[tb]
\epsfxsize=8.0cm
\epsfbox{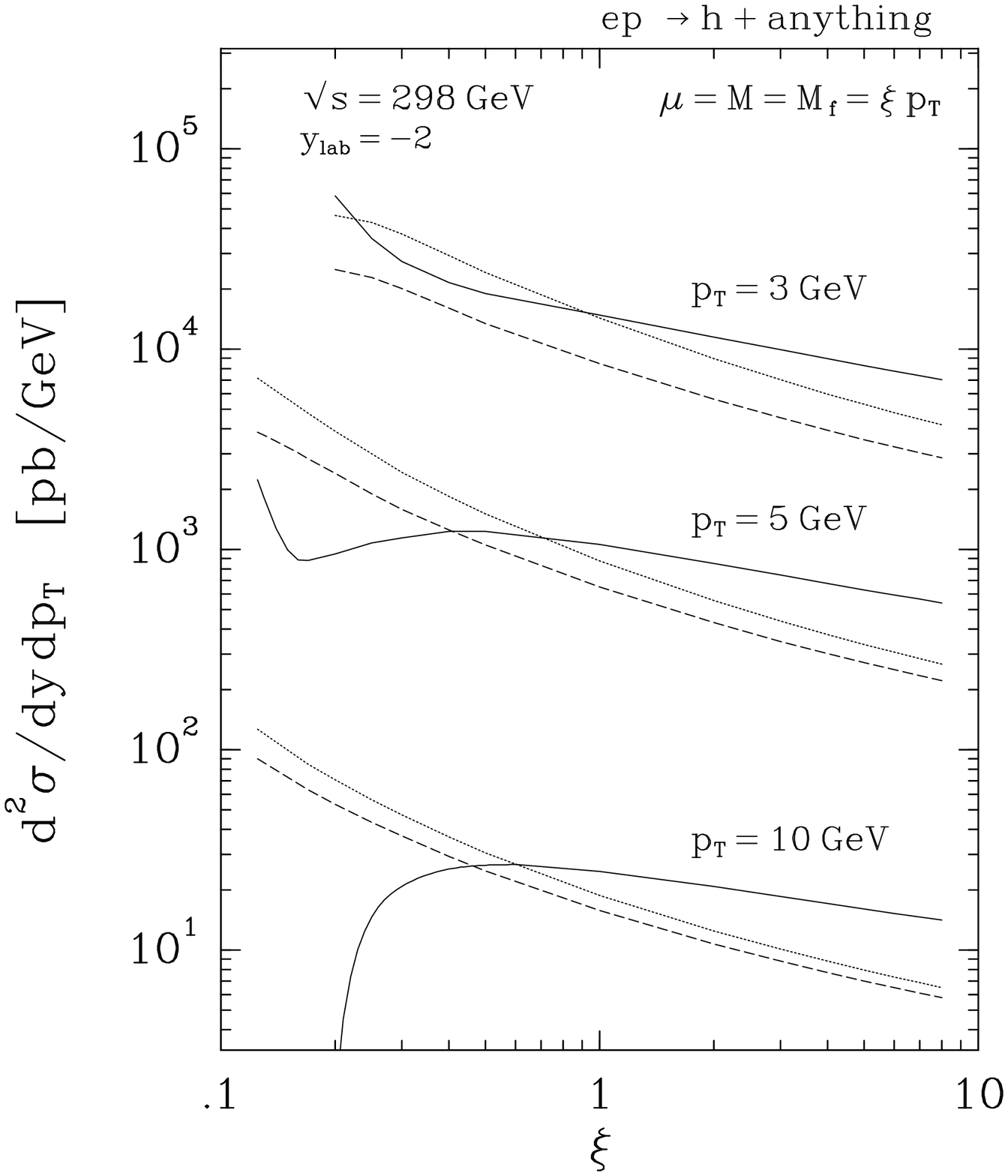}
{{\small {\bf Fig.~13.} Scale dependence of $d^2\sigma/dy\,d\pT$ for
the resolved contribution to $ep \to hX$ at $\sqrt{s}=298\,$GeV,
$\ylab=-2$, and three different values of $\pT$, $\pT=3,5,10\,$GeV.
The solid lines indicate the NLO result, the dotted lines the LO result,
and the dashed lines the LO result with
structure functions and $\as$ at the NLO level  } }
\end{figure}

To obtain more information on the scale dependence of the inclusive
cross section, we have calculated it as a function of the scale factor
$\xi$ defined in the previous sub-section for $\pT=3,5$ and $10\,$GeV and
$\sqrt{s}=298\,$GeV, $\ylab=-2$. The results are shown in
Fig.~13, where we compare the NLO predictions evaluated with
MT-B1 and GRV in \ms\ and two-loop $\as$ (solid lines) with the
LO predictions evaluated with MT-SL and GRV~(LO) and
one-loop $\as$ (dotted lines). To show the effect of the
$\O\left(\as^3\right)$ terms in the hard-scattering cross section, we
also plot the results obtained for the LO predictions when structure
function and $\as$ are kept at the NLO level.

We see quite clearly that the NLO cross section has less scale
dependence than the two LO ones. Except than for the case $\pT=3\,$GeV,
the NLO cross section develops a plateau for $\xi=0.6$.
The turn-over for small $\xi$ at
$\pT=10\,$GeV is presumably an artifact due to the use of LO
fragmentation functions. Both LO curves show a monotonic dependence on
$\xi$, which for $\xi>1$ is steeper than the dependence of the NLO cross
section. We also observe that the LO cross sections (dotted curves)
equal the NLO ones for $\xi$ between $0.6$ and $0.9$, whereas
at larger values of $\xi$ the two sets of curves are quite well
separated. These latter feature has also been observed in {\bf I} for
single-charged hadron production in $p \bar{p}$ collisions.

A comparison of Fig.~13 for $\pT=10\,$GeV with Fig.~6a, where the
scale dependence has been investigated in the case of
$\delta$-function fragmentation and real photoproduction at
$\sqrt{s}=256\,$GeV, yields the following observations. Apart from
the absolute size, the LO results exhibit roughly the same
qualitative features, which is attributed to a rather weak scale
dependence of the LO fragmentation functions. The situation
changes when NLO corrections are included. The drop-off at
small $\xi$ is stronger in Fig.~13 than in Fig.~6a.
Finally, higher scales are needed to obtain equal values of LO and NLO
results when realistic fragmentation functions are included.

\begin{figure}[tb]
\epsfxsize=8.0cm
\epsfbox{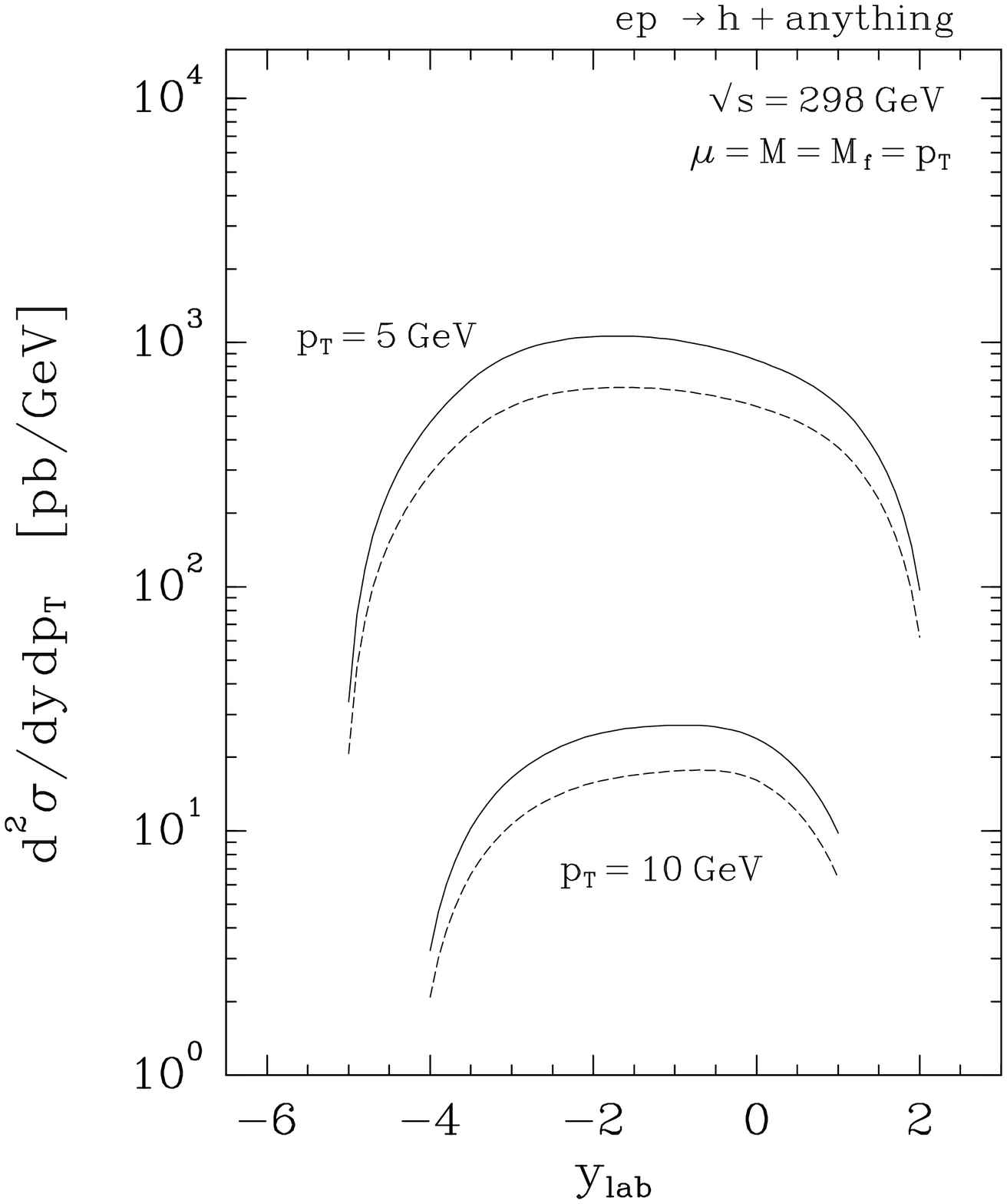}
{{\small {\bf Fig.~14.} Rapidity distribution of $d^2\sigma/dy\,d\pT$ for the
resolved contribution to $ep \to hX$ at $\sqrt{s}=298\,$GeV,
$\pT=5,10\,$GeV. The solid lines indicate the NLO result and the dashed
lines the LO result with structure functions and $\as$ at the NLO level}  }
\end{figure}

So far we have concentrated on the discussion of $\pT$ distributions
for fixed $\ylab=-2$, where cross sections are maximal. In Fig.~14 we
show the differential cross section $d^2\sigma/dy\,d\pT$ as a function
of the rapidity $y$ for $\sqrt{s}=298\,$GeV, $\mu=M=M_f=\pT$, and two
different values of $\pT$: $\pT=5$ and $10\,$GeV. We observe that these
distributions are quite symmetric around $\ylab \sim 2$ and that the
NLO corrections to the hard-scattering cross section do not change
their shape in any significant way.

\section{Summary and Conclusions}
We have presented next-to-leading-order predictions for
inclusive single-charged hadron and single-$\pi^0$ production
via resolved photoproduction at HERA.
We have confronted these predictions with first experimental data from
the H1 Collaboration and found that both are in good agreement
as for the slope in $\pT$.
To allow meaningful tests of the QCD-improved parton model and
to probe quantitatively the parton structure of the photon,
more accurate measurements are needed.

We find that the LO and NLO calculations coincide when the
renormalization and factorization scales are chosen to lie just
below $\pT$. We consider this agreement as accidental, since there
is a substantial discrepancy for other choices of scales. As
expected, the inclusion of NLO corrections leads to a reduction in
the scale dependence of the cross sections. Therefore, within the
scale variation in NLO, our results can be considered as an absolute
prediction for resolved photoproduction of single hadrons in $ep$
collisions. This is particularly important in the case of $\pi^0$
production, since such events provide a background for prompt-photon
emission. However, precisely in the case of neutral hadrons the
fragmentation functions used in our calculation may not be completely
reliable.

A further limitation of our analysis is related to the use of LO
fragmentation functions. Already now we observe that the dependence
on the fragmentation scale is reduced when the NLO terms of the
hard-scattering cross section are included. We expect the results to
become even more stable when also NLO fragmentation functions are utilized.

Finally, in order to disentangle experimentally resolved and direct
photoproduction, it is necessary to select small-$\pT$ hadrons
and/or to require the detection of the photon remnant jet. At present,
this does not represent a problem, since the data are still being
accumulated at relatively low $\pT$. However, as soon as data at
larger $\pT$ will become available, the direct process will have to be
evaluated in NLO and matched to the calculation presented here.

\noindent
{\bf Ackwnoledgments}

We thank L. Gordon for useful communications regarding [7].


{\footnotesize
\begin{itemize}
\item[1.]
  T. Ahmed et al., H1 Coll.: Phys.\ Lett.\ B297 (1992) 205
\item[{2.}]
  M. Derrick et al., ZEUS Coll.: Phys.\ Lett.\ B297 (1992) 404
\item[{3.}]
  J.G. Morfin, W.K. Tung: Z. Phys.\ C52 (1991) 13
\item[{4.}]
  A.D. Martin, R.G. Roberts, W.J. Stirling: Phys.\ Rev.\ D37 (1988) 1161;
     Mod.\ Phys.\ Lett.\ A4 (1989) 1135;
  P.N. Harriman, A.D. Martin, W.J. Stirling, R.G. Roberts:
     Phys.\ Rev.\ D42 (1990) 798;
  J. Kwiecinski, A.D. Martin, W.J. Stirling, R.G. Roberts:
     Phys.\ Rev.\ D42 (1990) 3645
\item[{5.}]
  A.D. Martin, W.J. Stirling, R.G. Roberts: Phys.\ Rev.\ D47 (1993) 867
\item[{6.}]
  M. Gl\"uck, E. Reya, A. Vogt: Phys.\ Rev.\ D45 (1992) 3986;
    Phys.\ Rev. D46 (1992) 1973
\item[{7.}]
  L.E. Gordon, J.K. Storrow: Z. Phys.\ C56 (1992) 307
\item[{8.}]
  P. Aurenche et al.: Z. Phys.\ C56 (1992) 589
\item[{9.}]
  P. Chiappetta et al.: Preprint CPT-92/\-PE.2841, $\qquad$
    ENSLAPP-\-A-\-416/93,
    FNT/T-92/46, IPNL 93-1 (December 1992)
\item[{10.}]
  R. Baier, J. Engels, B. Petersson: Z. Phys.\ C2 (1979) 265
\item[{11.}]
  M. Anselmino, P. Kroll, E. Leader, Z. Phys.\ C18 (1983) 307
\item[{12.}]
  F.M. Borzumati, B.A. Kniehl, G. Kramer:
    Z. Phys.\ C57 (1993) 595
\item[{13.}]
  F. Aversa, P. Chiappetta, M. Greco, J.Ph.~Guillet:
    Phys.\ Lett.\ B210 (1988) 225; ibid.\ B211 (1988) 465;
    Nucl.\ Phys.\ B327 (1989) 105
\item[{14.}]
  C.F. von Weizs\"acker, Z. Phys.\ 88 (1934) 612;
  E.J. Williams, Phys.\ Rev.\ 45 (1934) 729(L)
\item[{15.}]
  For a review see E. Paul: DESY Report 92--026 (February 1992)
\item[{16.}]
  H. Baer, J. Ohnemus, J.F. Owens: Z. Phys.\ C42 (1989) 657
\item[{17.}]
  K. Charchu\l a: Comput.\ Phys.\ Commun.\ 69 (1992) 360
\item[{18.}]
  M. Drees, K. Grassie: Z. Phys.\ C22 (1985) 451
\item[{19.}]
  P. Amaudruz et al., New Muon Coll.: Nucl.\ Phys.\ B371 (1992) 3;
    Phys.\ Lett.\ B295 (1992) 159;
  E. Kabuss: in: Proceedings of DESY-Zeuthen Workshop on Deep
    Inelastic Scattering, 1992
\item[{20.}]
  S.R. Mishra et al., CCFR Coll.:
    Columbia University Report, NEVIS \#1459 (June 1992)
\end{itemize}
 }
\end{document}